\documentclass[11pt]{article}
\usepackage[left=2.5cm, right=2.5cm,top=2.5cm,bottom=2.5cm,paper=a4paper]{geometry}
\usepackage[none]{hyphenat}
\usepackage{amsmath}
\usepackage{enumerate}
\usepackage{amsfonts}
\usepackage{xcolor}
\usepackage{array}
\usepackage{placeins}
\usepackage{multicol}
\usepackage{paracol}
\usepackage{amssymb}
\usepackage{stmaryrd}
\usepackage{changepage}
\usepackage{graphicx}
\usepackage{nicefrac, xfrac}
\usepackage{tikz}
\usetikzlibrary{shapes.arrows}

\begin{document}

\begin{center}
{\bf{\Large On Computability of Computable Problems\par}}
Asad Khaliq\\
E-mail address: asadmalik@giki.edu.pk\\
Faculty of Electrical Engineering, GIK Institute of Engineering Sciences and Technology.
KPK, Topi, 23640, Pakistan.\end{center}
%%%%%%%Abstract%%%%%%%%%%%%%
\vspace*{-4mm}
\begin{adjustwidth}{50pt}{50pt}  
{\bf Abstract:} Computational problems are classified into computable and uncomputable problems.If there exists an effective procedure (algorithm) to compute a problem then the problem is computable otherwise it is uncomputable.Turing machines can execute any algorithm therefore every computable problem is Turing computable.Cardinality of Turing machines and computable problems is equal-both are countably infinite.In this paper we introduce new type of problems by constructing a transform technique and applying it on some computable problems.The transformed problems can be computable of uncomputable.\\\vspace*{-5mm}\end{adjustwidth}

\section{Introduction}

Discovery of paradoxes during attempts to axiomatize Mathematics played important role in foundational crisis of Mathematics in late 19th and early 20th century.There were three major schools of thought to deal with this crisis, i.e., intuitionism, logisim, and formalism.Main opinions about this foundational crisis were turned into Mathematical research projects.David Hilbert, one of the most notable formalists, presented one such project that is known as Hilbert’s program \cite{ferreiros2008crisis}.Entscheidungsproblem, a decision problem of first order logic, was posed by David Hilbert and Wilhelm Ackermann in 1928 \cite{hilbert2022principles}.Entscheidungsproblem is about finding an effective procedure (algorithm) by which, given any expression $M$ in the notation of the system, it can be determined whether or not $M$ is provable in the system.Alan Turing and  Alonzo Church, in 1930’s, independently proved that Entscheidungsproblem is unsolvable \cite{bernays1936alonzo,turing1936computable}.To capture the notion of effective procedures, Alan Turing and Alonzo Church developed Turing machines and $\lambda $-calculus, respectively.General recursive functions and Post machines are some other formal notions of effective procedures that were also presented in 1930's \cite{godel1934undecidable,emil1936post}.Lambda calculus, Turing machines, and other notions of effective procedures are equivalent in their computational powers because they can simulate each other\cite{stephen1936kleene,COOPER2013145}.Only Turing machines are briefly described here because of their relevance to this work.

  A Turing machine consists of a finite set of states (control unit), a read-write head, and an infinite tape.A simple schematic of single tape Turing machine is shown in Fig. 1. At each step a Turing machine reads the alphabet from tape cell that is under read-write head, changes its state, writes an alphabet on the tape cell that is under read-write head, and moves the tape head to the left $(L)$ or to the right $(R)$.If $Q$ is finite set of states and $\Gamma$ represents finite set of tape alphabets then we can write transition function ($\delta$) of single tape Turing machine in following way.
  
\begin{equation*}
    \delta : Q \times \Gamma \longrightarrow Q \times \Gamma \times \{L,R\}
\end{equation*}

Unit time is required to execute one complete transition of transition function $\delta$.This unit time is same for every single transition of a transition function.If a problem is computable for given input then Turing machine halts in some finite time and outputs right answer.Now we construct a transform technique and apply it on some computable problems, and then we explain computability of the transformed problems.

 \newpage
%%%%%%%%%%%%%%%%%%%%Section 2%%%%%%%%%%%%%%%%%%%%%%%%%%%%%%%%%%%%%%%

\section{Construction of Transform}
\vspace*{-2mm}
~~~~Suppose there is a deterministic Turing machine $D_R$.The unit time that $D_R$ takes to complete one transition of its transition function is represented by $T_t$, and the time $T_t$ is constant and same for every Turing machine mentioned in this paper.Turing machine $D_R$ computes a partially computable function $R$.Suppose, for a given input $y$, the Turing machine $D_R$ requires $N_R$ number of steps to compute $R(y)$.If $R(y)$ is computable then $N_R$ is some finite number.For input $y$, the partially computable function $R$ can be either computable or uncomputable.Total time required to compute $R(y)$ by $D_R$ for input $y$ is represented by $f(R)$. 
\begin{equation}
    f(R) = N_R T_t
\end{equation}
With input $y$, there are two possible outcomes for partially computable function $R$, either $R(y)$ is a computable (if $D_R$ computes $R(y)$ and halts) or $R(y)$ is uncomputable (if $D_R$ loops forever).We can state these two possibilities in term of $f(R)$ in following way.For given input $y$
\vspace*{-1mm}
\begin{enumerate}[\hspace{-0.25 cm}I.]
\item If  $R(y)$ is computable then $f(R)$ has finite value 
\vspace*{-3mm}
\item If $f(R)$ has infinite value (i.e., $N_R \rightarrow \infty$ ) then $R(y)$ is uncomputable
\end{enumerate}
There is another a function $P$ and it is a total computable.If $P$ is total computable function then $f(P)$ has some finite value for every input.
\begin{equation}
    f(P) = N_P T_t
\end{equation}
   Suppose we have a variable $F$ such that $f(P,F) = N_P T_tF_P$ and $1\leq F_P<\infty$.If $M_P=\sfrac{1}{F_P}$ then $f(P,M) =\sfrac{N_P T_t}{M_P}$ and $0<M_P\leq 1$.This variable $M$ will be explained as we move forward in construction of this transform.The behaviour of (3), when $ M_P=1 $ and $ M_P\rightarrow0 $, is represented in (4) and (5), respectively.
\begin{equation}
    f(P, M) = \frac{N_P T_t}{M_P}
\end{equation}
\begin{equation}
    \lim_{{M_P = 1}} f(P, M) = f(P)
\end{equation}
\begin{equation}
    \lim_{{M_P \to 0}} f(P, M) \rightarrow \infty
\end{equation}
The new function $f(P,M)$ represents time required to compute function $P(m)$ for given input and $M$.Given that $P$ is computable function and $P(m)$ is new function, we can summarise (4) and (5) in following way.
\vspace*{1mm}

\hspace*{-6mm}For given input
\vspace*{-3mm}
\begin{enumerate}[\hspace{-0.25 cm}I.]
\item 	If $M_P=1$ then $f(P, M) = f(P)$; consequently, $P(m)$ is a computable function 
\vspace*{-3mm}
\item 	If $M_P\rightarrow0$ then $f(P, M) \rightarrow \infty$; consequently, $P(m)$ is an uncomputable function 
\end{enumerate}

In (3), if $M_P = 1 - 2^{-X_P}$ and $0<M_P\leq 1$ then $X_P = - \log_2(1-M_P)$ and $0<X_p\leq \infty $. After replacing $M_P$ with $1 - 2^{-X_P}$ in (3) we get (6).For any given input, $P$ is computable and computability of $P(x)$ depends on $x$, and in computation they require time $f(P)$ and $f(P,X)$, respectively.
\begin{equation}
    f(P, X) = \frac{N_P T_t}{1 - 2^{-X_P}}
\end{equation}
\begin{equation}
    \lim_{{X_P = \infty}} f(P,X) = f(P)
\end{equation}
\begin{equation}
    \lim_{{X_P \to 0}} f(P,X) \to \infty
\end{equation}
If it is given that $P$ is computable function but computability of $P(x)$ depends on value of $x$ then we can summarise (7 and 8) in following way.
\newpage

\hspace*{-6mm}For given input
\vspace*{-3mm}
\begin{enumerate}[\hspace{-0.25 cm}I.]
\item If $X_p=\infty$ then $f(P, X) = f(P)$; consequently, $P(x)$ is a computable function 
\vspace*{-3mm}

\item If $X_p\rightarrow0$ then $f(P, X) \to \infty$; consequently, $P(x)$ is an uncomputable function 
\end{enumerate}

\hspace{-7.5mm} We can implement partially computable and computable (partially computable and total) functions through Turing machines.For a given input, there always exists a Turing machine $M_1$ that can compute task $Y$ if there exists a computable function for $Y$.We can materialize the mathematical formulations presented in (6,7 and 8) through a realizable computational scheme that consists of Stipulation machine and Turing machine.Stipulation machine is a realizable computational object, just like Turing machines.Stipulation machine interacts with a Turing machine through some mutually binding set of rules (\emph{postulates}).

\vspace{-2mm}
%%%%%%%%%%%%%%%%%%%%%%%%%%%%%%%%%%%%%%%%%%%%%%%%%%%%%%%%%%%%%%%%
\subsection{Stipulation machine}
Suppose, there is a Turing machine $M_1$ that computes some problem $Y$, and $M_1$ is connected to Stipulation machine.A simple schematic of interconnection between Turing machine and Stipulation machine is shown in Fig.2. There are eight postulates that describe the interactions between Turing machine and Stipulation machine. \textbf{I.} Stipulation machine writes input on the space specified for input in Turing machine $M_1$, and there is one and only space for input in $M_1$. \textbf{II.} Stipulation machine writes only those inputs that $M_1$ can decide, i.e., $M_1$ is total Turing machine for the Stipulation machine. \textbf{III.}Stipulation machine rewrites input in Turing machine $M_1$ after every fixed time interval $T_Y$. \textbf{IV.}Length of every recurring input is equal to input length of first input in a computation. \textbf{V.} No two consecutive inputs of Turing machine $M_1$ can be identical in a computation. \textbf{VI.} When Turing machine $M_1$ halts the Stipulation machine gets this information. \textbf{VII.} When Stipulation machine receives halting signal, it stops rewriting input on Turing machine $M_1$ for current computation. \textbf{VIII.} We can write input in $M_1$ only through Stipulation machine when it is connected to Stipulation machine.
\vspace{1mm}

~~ The time $T_Y$ is characteristic of Stipulation machine and $0<T_Y\leq \infty$, and $T_Y=\infty$ means Stipulation machine does not rewrite input in a computation.If there is a Turing machine $M_1$ that is connected with Stipulation machine then it is represented by $M_1(t)$, and $\langle M_1(t)\rangle=\langle M_1\rangle$.Just like Turing machines, Stipulation machines is also a realizable object.Therefore, we can construct the computational scheme that is based on the eight postulates and interactions presented in Fig.2.It is important to note that Turing machine has only one place for input when it is connected to Stipulation machine and that place is accessible to Stipulation machine. 
\begin{figure}[h!]
    \centering

    \includegraphics[width=135mm, height=65mm]{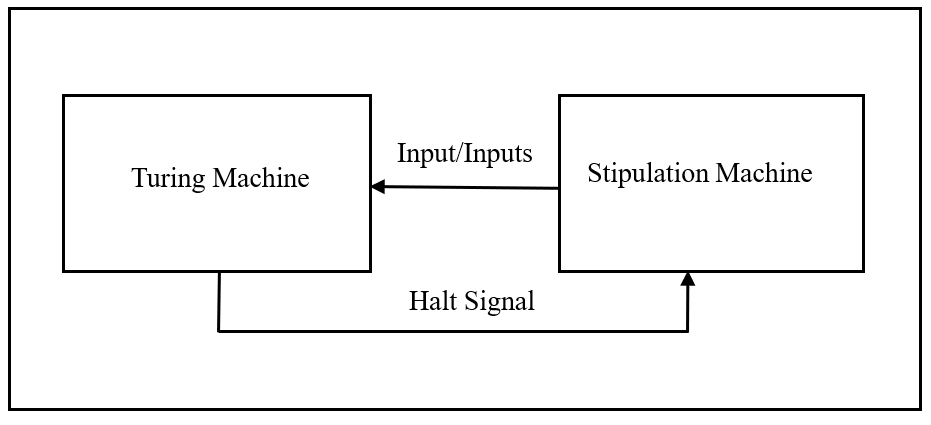}
    \caption{Flow of interactions between Turing
machine and Stipulation machine}
    \label{fig:enter-label}
\end{figure}
\vspace{-4mm}

\subsection{Computation by a Turing machine $P_r$ and $P_r(t)$}

~~~~There is a Turing machine $P_r$ that computes sum of input binary strings twice, it accepts if sums are \emph{\textbf{even and equal}} in both computations, it rejects if sums are \emph{\textbf{odd and equal}} in both computations, and it computes again from the start if sums in both computations is not identical (\emph{\textbf{even and equal}} or \emph{\textbf{odd and equal}}).Suppose there is set $A$ and $a \in A$, and $P_r$ decides for every input $a$.
\begin{equation*}
    A=\{110+101010, 1100+100, 1010+110, 100+10,1101110+10,\hdots  \}
\end{equation*}
\vspace{-2mm}
    
We give an input string $a$ (e.g., $1100+10110$) to $P_r$.The Turing machine $P_r$ decides and halts when given input is from $A$, i.e., $P_r(a)$ is decidable.Now, we transform $P_r$ into $P_r(t)$ by connected it to Stipulation machine and give input $a$ through Stipulation machine.In case of $P_r(t)$ with input from $A$, the behaviour of the decidability is quite different when $T_{P_r}$ is relatively large and $T_{P_r}$ is relatively small.In following two cases we analyse the decidability of $P_r(t)$ with input from $A$ and $0<T_{P_r}\leq \infty$; particularly, when $T_{P_r} \to 0$ and $T_{P_r}=\infty$.
\vspace{4mm}

\hspace{-5.5mm}\textbf{Case I.} Decidability of $P_{r}(t)$ with input $a$ and relatively large $T_{P_r}$.

\hspace{-7mm} Stipulation machine writes input string $a$ on Turing machine $P_r(t)$, the Turing machine $P_r(t)$ decides and halts on input $a$ in time less than $T_{P_r}$, Stipulation machines receives halts signal from $P_r(t)$ and does not rewrite input.If $T_{P_r}$ is relatively large then there is no difference between $P_r$ and $P_r(t)$, i.e., both  $P_r(a)$ and $P_r(t)(a)$ decide and halt.

With $T_{P_r}=\infty$ and input $a$, for all $a \in A$, both $P_r(a)$ and $P_r(t)(a)$ are decidable.There is no difference between $P_r$ and $P_r(t)$ when $T_{P_r}=\infty$.Therefore, $P_r(t)=P_r$ when $T_{P_r}=\infty$.
\vspace{4mm}

\hspace{-5.5mm}\textbf{Case II.} Decidability of $P_r(t)$ with input $a$ and $T_{P_r}\to 0 $.

\hspace{-7mm} If $T_{P_r}$ is relatively too small then inputs of $P_r(t)$ change too quickly.Now, the time $T_{P_r}$ is relatively so small that Stipulation machine rewrites input on $P_r(t)$ far before $P_r(t)$ decides and halts on an input.Consequently, $P_r(t)$ carries out computation on \emph{\textbf{some parts}} of each recurring input in \emph{\textbf{some order}}.The transitions that $P_r(t)$ takes depends on its transition function, current configuration and those \emph{\textbf{some parts}} of recurring inputs.Those \emph{\textbf{some parts}} and \emph{\textbf{some order}} make $P_r(t)$ to take transitions that it would take if input is some unknown string $f$, and length and composition of $f$ depends on $T_{P_r}$.The arbitrary string $f$ may not be element of set $A$.
\begin{equation*}
    L_2=\{ \langle P_r(t),f\rangle| \text{ Turing machine } P_r(t)  \text{ decides } f\}
\end{equation*}

~~~~The language $L_2$ is undecidable because decidability of a Turing machine on an arbitrary string is undecidable.
So, the decidability of $P_r(t)(a)$ depends on $T_{P_r}$. If $T_{P_r}$ is relatively large then $P_r(t)(a)$ is decidable but if $T_{P_r}$ is too small (i.e., $T_{P_r}\to 0$)  then $P_r(t)(a)$ is undecidable.
\vspace{2mm}

\hspace{-6mm}Only end points of $T_{P_r}$ (i.e., $T_{P_r}\to 0$ and $T_{P_r}= \infty$) are considered here but $T_{P_r}$ can have any value on interval $(0  \quad \infty]$ and it is discussed in next section (Section 3).The time $P_r$ takes to decide $a$ in $P_r(a)$ is represented by $f(P_r)$, and the time $P_r(t)$ takes to recognize $a$ in $P_r(t)(a)$ is represented by $f(P_r,T)$.Following mathematical expression can represent the conclusions of case I and II.

\begin{equation}
    f(P_r, T) = \frac{N_{P_r} T_t}{1 - 2^{-T_{P_r}}}
\end{equation}
\vspace*{-4mm}
\begin{equation}
    \lim_{{T_{P_r} = \infty}} f(P_r, T) = f(P_r)
\end{equation}
\begin{equation}
    \lim_{{T_{P_r} \to 0}} f(P_r, T) \to \infty
\end{equation}
\newpage
Given that $P_r(a)$ is decidable
\vspace*{-2mm}
\begin{enumerate}[\hspace{-0.25 cm}I.]
\item 			If $T_{P_r }=\infty$ then $f(P_r, T) = f(P_r)$; consequently,   $P_r(t)(a)$ is decidable 
\vspace*{-3mm}
\item 			If $T_{P_r }\rightarrow0$ then  $f(P_r, T) \to \infty$; consequently,   $P_r(t)(a)$ is undecidable 
\end{enumerate}

Stipulation machine and Turing machine are realizable computational objects.Therefore, Stipulation machines and Turing machines can be designed to experimentally validate this  transform.Now we construct and describe some examples where a problem $P(\psi)$ is computable for input $\psi$, and we transform $P(\psi)$ into $P(t)(\psi)$.The problem $P$ can be a decision or a function problem.
\vspace*{-3mm}

\subsection{Examples of Transformed Problems}
\vspace*{-2mm}

The problem $P_{1}$ is to evaluate a 3-SAT (e.g., 12) twice 
 for a given input $n_{\Phi}$, if both answers are yes then accept, if both answers are no then  reject; and if both answers are not same then start evaluating again and continue until we have two consecutive same answers.If $P_{1}$ is decidable for given input $n_{\Phi}$ then it will remain decidable if we compute it for any number of times instead of two times-because we will have same answer (yes or no) irrespective how many time we evaluate a formula on a given input.We transform $P_1$ into $P_1(t)$, and variation time of Stipulation machine for $P_{1}(t)$ is represented by $T_{P_{1}}$.In computation of $P_{1}(t)(n_{\Phi})$, we give input $n_{\Phi}$ to the Turing machine  through stipulation machine.
\begin{equation}
   \Phi = (x_1 \lor \overline{x_5} \lor x_7) \land (\overline{x_2} \lor \overline{x_4} \lor x_8) \land (\overline{x_2} \lor \overline{x_3} \lor x_6)
\end{equation}
\begin{equation}
    \Phi_d = (x_1(t) \lor \overline{x_5(t)} \lor x_7(t)) \land (\overline{x_2(t)} \lor \overline{x_4(t)} \lor x_8(t)) \land (\overline{x_2(t)} \lor \overline{x_3(t)} \lor x_6(t))
\end{equation}
 
\begin{equation}
    f(P_{1}, T) = \frac{N_{P_{1}} T_t}{1 - 2^{-T_{P_{1}}}}
\end{equation}
\vspace*{-7mm}

\begin{equation}
    \lim_{{T_{P_{1}} = \infty}} f(P_{1}, T) = f(P_{1})
\end{equation}
\begin{equation}
    \lim_{{T_{P_{1}} \to 0}} f(P_{1}, T) \to \infty
\end{equation}
\vspace*{-2mm}

\hspace{-5mm}Given that ${P_{1}}(n_{\Phi})$ is decidable 
\begin{enumerate}[\hspace{-0.25 cm}I.]
\item 		If $T_{P_{1}} =\infty$ then $f(P_{1}, T) = f(P_{1})$; consequently, $P_1(t)(n_{\Phi})$ is decidable problem  
\vspace*{-3mm}
\item 		If $T_{P_{1}} \rightarrow0$ then $f(P_{1}, T) \to \infty$; consequently, $P_1(t)(n_{\Phi})$ is undecidable problem
\end{enumerate}

Even when $P_1(t)(n_{\Phi})$ is decidable for every recurring input $n_{\Phi}$ but $P_1(t)$ may never halt if $T_{P_{1}}$ is too small.Through this transform technique, we can map a decidable decision problem into a decidable problem or an undecidable problem.Suppose there are two $n\times n$ non-zero matrices $A$, and $B$ such that $AB=C$.The problem $P_{2}$ is to compute $C$ four times for given $A$ and $B$, if in all four answers $c_{11}c_{1n} \geq c_{nn}c_{n1}$ then accept, if in all four answers $c_{11}c_{1n} < c_{nn}c_{n1}$ then reject; and if all four answers are not same then compute $C$ again from start and continue until we have four consecutive same answers.We transform $P_2$ into $P_2(t)$, and variation time of Stipulation machine for $P_{2}(t)$ is represented by $T_{P_{2}}$.In computation of $P_2(t)(A,B)$, we give input ($A$,$B$) to the Turing machine through stipulation machine.Decidability of transformed problem $P_2(t)(A,B)$ depends on $T_{P_{2}}$.
%%%%%%%%%%%%%%parallele two matrix%%%%%%%%%%%%%%%%%%%%%%%%%

\begin{equation*}
     \begin{bmatrix}
    a_{11} & a_{12} & a_{13} & \cdots & a_{1n} \\
    a_{21} & a_{22} & a_{23} & \cdots & a_{2n} \\
    a_{31} & a_{32} & a_{33} & \cdots & a_{3n} \\
    \vdots & \vdots & \vdots & \ddots & \vdots \\
    a_{n1} & a_{n2} & a_{n3} & \cdots & a_{nn}
\end{bmatrix}\times
\begin{bmatrix}
    b_{11} & b_{12} & b_{13} & \cdots & b_{1n} \\
    b_{21} & b_{22} & b_{23} & \cdots & b_{2n} \\
    b_{31} & b_{32} & b_{33} & \cdots & b_{3n} \\
    \vdots & \vdots & \vdots & \ddots & \vdots \\
    b_{n1} & b_{n2} & b_{n3} & \cdots & b_{nn} 
\end{bmatrix}
=
 \begin{bmatrix}
    c_{11} & c_{12} & c_{13} & \cdots & c_{1n} \\
    c_{21} & c_{22} & c_{23} & \cdots & c_{2n} \\
    c_{31} & c_{32} & c_{33} & \cdots & c_{3n} \\
    \vdots & \vdots & \vdots & \ddots & \vdots \\
    c_{n1} & c_{n2} & c_{n3} & \cdots & c_{nn} 
\end{bmatrix}
\end{equation*}
\vspace{3mm}
\begin{equation}
    f(P_2, T) = \frac{N_{P_2} T_t}{1 - 2^{-T_{P_2}}}
\end{equation}
\begin{equation}
    \lim_{{T_{P_2}= \infty}} f(P_2, T) = f(P_2)
\end{equation}
\begin{equation}
    \lim_{{T_{P_2} \to 0}} f(P_2, T) \to \infty
\end{equation}

\hspace{-5mm}Given that $P_2(A,B)$ decidable 
\begin{enumerate}[\hspace{-0.25 cm}I.]
\item 			If $T_{P_{2}} =\infty$ then $f(P_{2}, T) = f(P_{2})$; consequently, $P_2(t)(A,B)$ is decidable
\vspace*{-3mm}
\item 			If $T_{P_{3}} \to 0$ then $f(P_{3},T) \to \infty$; consequently, $P_3(t)(A,B)$ is undecidable
\end{enumerate}

Suppose there are two $n\times n$ matrices $A$ and $B$, and both matrices have integer entries.The problem $P_{3}$ is to compute $AB$ thrice, and in all three computations, if $AB=I$ then accept, if $AB\neq I$ then reject; and if all three answers are not same then compute $AB$ again from start and continue until we have three consecutive same answers.We transform $P_3$ into $P_3(t)$, and variation time of Stipulation machine for $P_{3}(t)$ is represented by $T_{P_{3}}$.In computation of $P_3(t)(A,B)$, we give input ($A$, $B$) to the Turing machine through stipulation machine.Decidability of $P_3(t)(A,B)$ depends on variation time $T_{P_{3}}$.
%%%%%%%%%%%%%%parallele two matrix%%%%%%%%%%%%%%%%%%%%%%%%%
\vspace*{-3mm}
\begin{paracol}{2}
\begin{equation*}
     A=\begin{bmatrix}
    a_{11} & a_{12} & a_{13} & \cdots & a_{1n} \\
    a_{21} & a_{22} & a_{23} & \cdots & a_{2n} \\
    a_{31} & a_{32} & a_{33} & \cdots & a_{3n} \\
    \vdots & \vdots & \vdots & \ddots & \vdots \\
    a_{n1} & a_{n2} & a_{n3} & \cdots & a_{nn}
\end{bmatrix}
\end{equation*}
\switchcolumn
\begin{equation*}
   B = \begin{bmatrix}
    b_{11} & b_{12} & b_{13} & \cdots & b_{1n} \\
    b_{21} & b_{22} & b_{23} & \cdots & b_{2n} \\
    b_{31} & b_{32} & b_{33} & \cdots & b_{3n} \\
    \vdots & \vdots & \vdots & \ddots & \vdots \\
    b_{n1} & b_{n2} & b_{n3} & \cdots & b_{nn} 
\end{bmatrix}
\end{equation*}
\end{paracol}

\begin{equation*}
     \begin{bmatrix}
    a_{11} & a_{12} & a_{13} & \cdots & a_{1n} \\
    a_{21} & a_{22} & a_{23} & \cdots & a_{2n} \\
    a_{31} & a_{32} & a_{33} & \cdots & a_{3n} \\
    \vdots & \vdots & \vdots & \ddots & \vdots \\
    a_{n1} & a_{n2} & a_{n3} & \cdots & a_{nn}
\end{bmatrix}\times
\begin{bmatrix}
    b_{11} & b_{12} & b_{13} & \cdots & b_{1n} \\
    b_{21} & b_{22} & b_{23} & \cdots & b_{2n} \\
    b_{31} & b_{32} & b_{33} & \cdots & b_{3n} \\
    \vdots & \vdots & \vdots & \ddots & \vdots \\
    b_{n1} & b_{n2} & b_{n3} & \cdots & b_{nn} 
\end{bmatrix}
=
 \begin{bmatrix}
    1 & 0 & 0 & \cdots & 0 \\
    0 & 1 & 0 & \cdots & 0 \\
    0 & 0 & 1 & \cdots & 0 \\
    \vdots & \vdots & \vdots & \ddots & \vdots \\
    0 & 0 & 0 & \cdots & 1 
\end{bmatrix}
\end{equation*}
\

\begin{equation}
    f(P_3, T) = \frac{N_{P_3} T_t}{1 - 2^{-T_{P_3}}}
\end{equation}

\begin{equation}
    \lim_{{T_{P_3} = \infty}} f(P_3, T) = f(P_3)
\end{equation}
\begin{equation}
    \lim_{{T_{P_3} \to 0}} f(P_3, T) \to \infty
\end{equation}

\hspace{-6mm} Given that $P_3(A,B)$ is decidable 
\begin{enumerate}[\hspace{-0.25 cm}I.]
\item 			If $T_{P_{3}} =\infty$ then $f(P_{3}, T) = f(P_{3})$; consequently, $P_3(t)(A,B)$ is decidable
\vspace*{-3mm}
\item 			If $T_{P_{2}} \to 0$ then $f(P_{2},T) \to \infty$; consequently, $P_3(t)(A,B)$ is undecidable
\end{enumerate}
\vspace*{-1.5mm}

So far, we have applied this transform technique on decidable decision problems, but we can also apply it on computable functions problems.Suppose  there is some binary string $u$.The problem $P_4$ is to compute the number of ones in $u$ for $K$ times for some $K \in \mathbb{N}$, if number of ones are equal in each count then output the result and halt, and if number of ones in each count are not equal then start counting again and continue until we have equal number of ones in all $K$ counts and output the count.We transform $P_4$ into $P_4(t)$, and variation time of Stipulation machine for $P_{4}(t)$ is represented by $T_{P_{4}}$.Given that the problem $P_4(u)$ is computable, and we give input $u$ to $P_{4}(t)$ through Stipulation machine for computation of $P_{4}(t)(u)$.Just like in any other transformed problem the value of time $T_{P_{4}}$ plays decisive role in computability of $P_{4}(t)(u)$.

\begin{equation}
    f(P_4, T) = \frac{N_{P_4} T_t}{1 - 2^{-T_{P_4}}}
\end{equation}
\begin{equation}
    \lim_{{T_{P_4} = \infty}} f(P_4, T) = f(P_4)
\end{equation}
\begin{equation}
    \lim_{{T_{P_4} \to 0}} f(P_4, T) \to \infty
\end{equation}

\hspace{-6mm} Given that $P_4(u)$ computable 
\begin{enumerate}[\hspace{-0.25 cm}I.]
\item 			If $T_{P_{4}} =\infty$ then $f(P_{4}, T) = f(P_{4})$; consequently, $P_4(t)(u)$ is computable
\vspace*{-3mm}
\item 			If $T_{P_{4}} \to 0$ then $f(P_{4},T) \to \infty$; consequently, $P_4(t)(u)$ is uncomputable
\end{enumerate}
 Suppose there is an $m\times n$ matrix A with $a_{ij}\in \mathbb{N}$, and A is invertible matrix.The problem $P_5$ is to compute $A^{-1}$, prove $AA^{-1}=I$ and halt, and if $AA^{-1} \neq I$ then start again and continue until $AA^{-1}=I$ for some invertible matrix $A$.The problem $P_5$ is a computable when input is $A$.We transform $P_5$ into $P_5(t)$, and variation time of Stipulation machine for $P_{5}(t)$ is represented by $T_{P_{5}}$.For input $A$ the problem $P_5(A)$ is computable, and we give $A$ to $P_{5}(t)$ through Stipulation machine in computation of $P_{5}(t)(A)$.

\begin{equation}
    A = \left[ \begin{array}{cccccc}
    a_{11} & a_{12} & a_{13} & \cdots & a_{1n} \\
    a_{21} & a_{22} & a_{23} & \cdots & a_{2n} \\
    a_{31} & a_{32} & a_{33} & \cdots & a_{3n} \\
    \vdots & \vdots & \vdots & \ddots & \vdots \\
    a_{m1} & a_{m2} & a_{m3} & \cdots & a_{mn}
    \end{array} \right]
\end{equation}

\begin{equation}
    A(t) = \left [\begin{array}{cccccc}
    a_{11}(t) & a_{12}(t) & a_{13}(t) & \cdots & a_{1n}(t) \\
    a_{21}(t) & a_{22}(t) & a_{23}(t) & \cdots & a_{2n}(t) \\
    a_{31}(t) & a_{32}(t) & a_{33}(t) & \cdots & a_{3n}(t) \\
    \vdots & \vdots & \vdots & \ddots & \vdots \\
    a_{m1}(t) & a_{m2}(t) & a_{m3}(t) & \cdots & a_{mn}(t)
\end{array} \right]
\end{equation}
\begin{equation}
    f(P_5, T) = \frac{N_{P_5} T_t}{1 - 2^{-T_{P_5}}}
\end{equation}
\begin{equation}
    \lim_{{T_{P_5} = \infty}} f(P_5, T) = f(P_5)
\end{equation}
\begin{equation}
    \lim_{{T_{P_5} \to 0}} f(P_5, T) \to \infty
\end{equation}

\hspace{-4mm} Given that $P_5(A)$ computable 
\begin{enumerate}[\hspace{-0.25 cm}I.]
\item 			If $T_{P_{5}} =\infty$ then $f(P_{5}, T) = f(P_{5})$; consequently, $P_5(t)(A)$ is computable
\vspace*{-3mm}
\item 			If $T_{P_{5}} \to 0$ then $f(P_{5},T) \to \infty$; consequently, $P_5(t)(A)$ is uncomputable
\end{enumerate}

Suppose there is an $n\times n$ matrix $Q$ with $q_{ij}\in \mathbb{N}$, and rank of $Q$ is $n$.The problem $P_6$ is to compute inverse of input matrix and then inverse of inverse matrix, if double inverse matrix is identical to input matrix then halt otherwise compute this task again from start on input matrix.The problem $P_6$ is computable if input is $Q$ because  $(Q^{-1})^{-1}=Q$.We transform $P_6$ into $P_6(t)$, and variation time of Stipulation machine for $P_{6}(t)$ is represented by $T_{P_{6}}$.There is $Q$ for which $P_6(Q)$ is computable, and then we give $Q$ to $P_{6}(t)$ through Stipulation machine for computation of $P_6(t)(Q)$.

\begin{equation*}
     Q=\begin{bmatrix}
    q_{11} & q_{12} & q_{13} & \cdots & q_{1n} \\
    q_{21} & q_{22} & q_{23} & \cdots & q_{2n} \\
    q_{31} & q_{32} & q_{33} & \cdots & q_{3n} \\
    \vdots & \vdots & \vdots & \ddots & \vdots \\
    q_{n1} & q_{n2} & q_{n3} & \cdots & q_{nn}
\end{bmatrix}
\end{equation*}
\vspace{2mm}

\hspace{-6mm} Given that $P_6(Q)$ computable 
\begin{enumerate}[\hspace{-0.25 cm}I.]
\item 			If $T_{P_{6}} =\infty$ then $f(P_{6}, T) = f(P_{6})$; consequently, $P_6(t)(Q)$ is computable
\vspace*{-3mm}
\item 			If $T_{P_{6}} \to 0$ then $f(P_{6},T) \to \infty$; consequently, $P_6(t)(Q)$ is uncomputable
\end{enumerate}

Suppose there is a problem $P(\psi)$ and it is  computable with input $\psi$.We transform this computable problem $P(\psi)$ into $P(t)(\psi)$ through a transform that will be represented by $f(Z,T)$; and transformed problem $P(t)(\psi)$ can be computable or uncomputable, depending on time $T_P$.\\

\begin{tikzpicture}[every path/.style={line width=1pt}]

  % Draw the initial node
  \node (start) at (0,0) {$P(\psi)$ (Computable)};
  % Draw the arrow with a+b split into two lines
  \draw(start) -- node[above] {$f(Z,T) \hspace{0.5em} Transform$} ++(5.5,0);
 \draw [->, >=stealth](5.5,0) parabola  (7.5,1.5) node[ right]{ $P(t)(\psi)$ (Computable if $T_P= \infty$)};
 \draw [->, >=stealth](5.5,0) parabola  (7.5,-1.5)node[ right]{ $P(t)(\psi)$ (Uncomputable if $T_P \to 0$)};
\end{tikzpicture}

\hspace{-7.5mm} The time $T_P \in (0 \text{  }  \infty]$ but we have discussed only two cases, i.e., $T_P = \infty$ and $T_P \to 0$. Stipulation machine can have any value of $T_P$ on interval $(0 \text{  }  \infty]$.Now we discuss the behaviour of $f(Z,T)$ transform when $T_P$ can have any possible value on interval $(0 \text{  }  \infty]$ for transformed problem $P(t)(\psi)$.

\begin{tikzpicture}[every path/.style={line width=1pt}]

  % Draw the initial node
  \node (start) at (0,0) {$P(\psi)$ (Computable)};
  % Draw the arrow with a+b split into two lines
  \draw(start) -- node[above] {$f(Z,T) \hspace{0.5em} Transform$} ++(5,0);
 \draw [->, >=stealth](5,0) parabola  (7,1.5) node[right]{ $P(t)(\psi)$ (Computable if $T_P= \infty$)};
  \draw [->, >=stealth](5,0) parabola  (7,1.2);
  \draw [->, >=stealth](5,0) parabola  (7,0.9);
  \draw [->, >=stealth](5,0) parabola  (7,0.6)node[above right]{\vdots};
  \draw [->, >=stealth](5,0) parabola  (7,0.3);
  \draw [->, >=stealth](5,0) parabola  (7,0);
 %lowe half
 \draw [->, >=stealth](5,0) parabola  (7,-1.5)node[right]{ $P(t)(\psi)$ (Uncomputable if $T_P \to 0$)};
 \draw [->, >=stealth](5,0) parabola  (7,-1.2);
  \draw [->, >=stealth](5,0) parabola  (7,-0.9)node[above right]{\vdots};
  \draw [->, >=stealth](5,0) parabola  (7,-0.6);
  \draw [->, >=stealth](5,0) parabola  (7,-0.3)node[above right]{ $P(t)(\psi)$ (If $T_P\in (0 \quad \infty)$ then?)};

\end{tikzpicture}

\vspace*{-2mm}
\section{ Normalization of $f(Z,T)$ Transform}
\vspace*{-2mm}
Suppose there is problem $P$ that is computable with input $\psi$, and $\psi \in \Psi$ and $L(P)=\Psi$.We apply $f(Z,T)$ transform on $P(\psi)$ and get $P(t)(\psi)$.Normalization of the transformed problem $P(t)(\psi)$ is defined in following way.
\begin{equation}
    f(P, T) = \frac{N_P T_t}{1 - 2^{-T_P}}
\end{equation}
\begin{equation}
  f_n(P, T) = \frac{f(P, T)}{\frac{T_P}{1 - 2^{-T_P}}}
\end{equation}
\begin{equation}
    f_n(P, T) = \frac{N_P T_t}{T_P}
\end{equation}
Because $P(\psi)$ is computable problem and $0<T_P\leq \infty$ therefore $0\leq f_n (P,T)<\infty$.The problem $P(t)(\psi)$ is computable if $0 \leq f_n (P,T)<1$ because $P(t)(\psi)$ halts before Stipulation machine rewrites input. If $N_P T_t>T_P$ then $f_n (P,T)>1$ i.e., input of $P(t)$ varies before $P(t)(\psi)$ gets computed and $P(t)$ halts.According to postulate \textbf{V}, no two consecutive inputs of $P(t)$ can identical in a computation.If $f_n (P,T)>1$ then Stipulation machine rewrites input on $P(t)$ before $P(t)$ decides and halts on input $\psi$.Consequently, $P(t)$ carries out computation on \emph{\textbf{some parts}} of each recurring input in \emph{\textbf{some order}}.Those \emph{\textbf{some parts}} and \emph{\textbf{some order}} make arbitrary string $\psi_f$ of some arbitrary length, and composition of $\psi_f$ depends on $T_{P}$.The arbitrary string $\psi_f$ may not be in set $\Psi$.Computation of $P(t)$ on an arbitrary string $\psi_f$ is uncomputable.Therefore, if $f_n (P,T)>1$ then $P(t)(\psi)$ is not computable but if $0\leq f_n (P,T)<1$ then $P(t)(\psi)$ is computable.Normalization line for $P(t)(\psi)$ is shown in Fig.3.

\begin{figure}[h!]
    \centering
    \includegraphics[width=135mm, height=57mm]{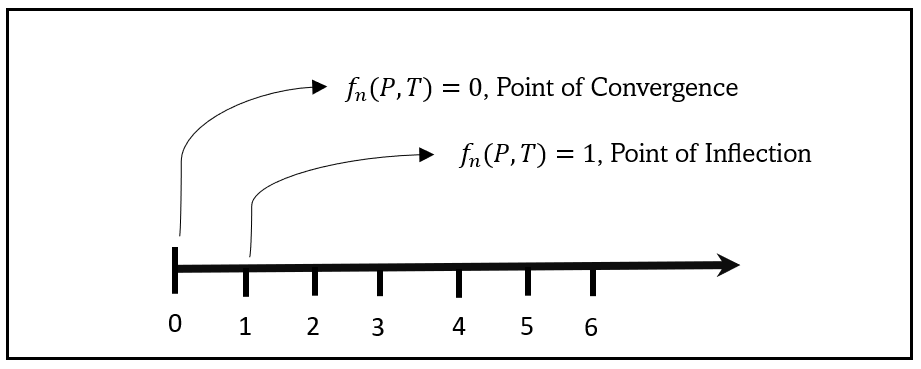}
    \caption{Normalization line of $f(Z,T)$ Transform}
    \label{fig:enter-label}
\end{figure}
\FloatBarrier
\vspace*{-3mm}
Point of inflection and point of convergence are two important points on normalization line. If $f_n(P,T)=0$ then transformed problem $P(t)(\psi)$ exist at point of convergence and $f(P,T)=f(P)$ when $f_n (P,T)=0$.If $f_n (P,T)=1$ then transformed problem $P(t)(\psi)$ exist at point of inflection.The point where $f_n (P,T)=1$ is very critical point for computability of problem $P(t)(\psi)$.A small change can make $P(t)(\psi)$ move in-between computable-uncomputable when $f_{n}(P,T)$ exists around point of inflection.If $P(\psi)$  requires $N_P$ steps and $P(t)(\psi)$ exists around point of inflection, then slight change in $N_P$ (e.g., $N_{P}-n$ or $N_P+n$ for some $n \geq 1$) or $T_P$ can change the computability of $P(t)(\psi)$.In computation of $P(t)(\psi)$, if $f_n (P,T)<1$ then Turing machine carries out computation on input string $(\psi)$ but if $f_n (P,T)>1$ then Turing machine carries out computation on some arbitrary string that is composed of some input strings from $\Psi$.We can also represent normalization line of $P(t)(\psi)$ in following way.
\vspace{2mm}

\begin{tikzpicture}[every path/.style={line width=1pt}]

  % Draw the initial node
  \node (start) at (0,0) {$P(\psi)$ (Computable)};
  % Draw the arrow with a+b split into two lines
  \draw(start) -- node[above] {$f(Z,T) \hspace{0.5em} Transform$} ++(6,0);
 \draw [->, >=stealth](5,0) parabola  (7.2,1.5) node[above right]{ $P(t)(\psi)$ (If $T_P= \infty$ then $f_n(P,T)=0$)};
  \draw [->, >=stealth](5,0) parabola  (7.2,1.2);
  \draw [->, >=stealth](5,0) parabola  (7.2,0.9)node[above right]{ $P(t)(\psi)$ ($0<f_n(P,T)<1$)};
  \draw [->, >=stealth](5,0) parabola  (7.2,0.6);
  \draw [->, >=stealth](5,0) parabola  (7.2,0.3)node[above right]{\vdots};
  \draw [->, >=stealth](5,0) parabola  (7.2,0);
 %lowe half
 \draw [->, >=stealth](5,0) parabola  (7,-1.5)node[below right]{ $P(t)(\psi)$ (If $T_P \to 0$ then $f_n(P,T)\to \infty$)};
 \draw [->, >=stealth](5,0) parabola  (7.2,-1.2)node[right]{\vdots};
  \draw [->, >=stealth](5,0) parabola  (7.2,-0.9);
  \draw [->, >=stealth](5,0) parabola  (7.2,-0.9)node[above right]{ $P(t)(\psi)$ ($f_n(P,T)>1$)};
  \draw [->, >=stealth](5,0) parabola  (7.2,-0.6);
  \draw [->, >=stealth](5,0) parabola  (7.2,-0.3)node[above right]{ $P(t)(\psi)$ ($f_n(P,T)= 1$ for some $T_P$)};

\end{tikzpicture}
\vspace{-2mm}

The $f(Z,T)$ transformed problems that exist just before or immediately after point of inflection are very helpful in analysing the computational characteristics of different variants of Turing machines.If cardinality of all $f(Z,T)$ transformed problems is higher than cardinality of Turing machines then there must exist $f(Z,T)$ transformed problems that are uncomputable.
\vspace{-4mm}

\section{Cardinality of $f(Z,T)$ Transformed Problems}
\vspace{-1mm}

The $f(Z,T)$ transform is applied to only computable function (or decidable languages) and it produces transformed problems.Suppose there is a language $L_C$ and it has only finite number of strings in it, and $l_C \in L_C$.There is a Turing machine $P_C$ that can decide $L_C$, i.e., $L(P_C)=L_C$ and $P_C$ halts on all inputs.We want to know the cardinality of $f(Z,T)$ transformed problems if we apply $f(Z,T)$ transform on $L_C$.We change $P_C$ into $P_C(t)$ and input is from $L_C$ through stipulation machine. If $0<T_{P_C} \leq \infty$ and then $|T_{P_C}|=\aleph_0$, and if $|T_{P_C}|=\aleph_0$ then $f(Z,T)$ transform produces infinite number of transformed problems in the form of $P_C(t)(l_C)$ for $T_{P_C}(1)$, $P_C(t)(l_C)$ for $T_{P_C}(2)$, $P_C(t)(l_C)$ for $T_{P_C}(3)$, $P_C(t)(l_C)$ for $T_{P_C}(4)$,  \ldots $P_C(t)(l_C)$ for $T_{P_C}(\infty)$, and $T_{P_C}=\{T_{P_C}(1),\text{ } T_{P_C}(2),\text{ } T_{P_C}(3),\text{ } \ldots,\text{ }T_{P_C}(\infty) \}$.The cardinality of $f(Z,T)$ transformed problems is countably infinite when we apply $f(Z,T)$ transform on a decidable language that has only finite number of strings in it, e.g., language $L_C$.
\begin{equation*}
    L_C=\{ l_{C(1)},\text{ } l_{C(2)},\text{ } l_{C(3)},\text{ } l_{C(4)},\text{ } l_{C(5)}, \text{ } \hdots,\text{ }  l_{C(128)}\}
\end{equation*}

     \begin{tikzpicture}[every path/.style={line width=1pt}]

  % Draw the initial node
  \node (start) at (0,0) {$P_C(l_C)$ (Decidable)};
  % Draw the arrow with a+b split into two lines
  \draw(start) -- node[above] {$f(Z,T) \hspace{0.5em} Transform$} ++(6,0);
 \draw [->, >=stealth](6,0) parabola  (8,2.5) node[right]{ $P_C(t)(l_C)$ for $T_{P_C}(1)$};
\draw [->, >=stealth](6,0) parabola  (8.2,2.0)node[right]{ $P_C(t)(l_C)$ for $T_{P_C}(2)$};
\draw [->, >=stealth](6,0) parabola  (8.2,1.5)node[right]{ $P_C(t)(l_C)$  for $T_{P_C}(3)$};
\draw [->, >=stealth](6,0) parabola  (8.2,1)node[ above right]{\vdots};
\draw [->, >=stealth](6,0) parabola  (8.2,0.5)node[above right]{\vdots};
\draw [->, >=stealth](6,0) parabola  (8.2,0)node[above right]{\vdots};
%lowe half
 \draw [->, >=stealth](6,0) parabola  (8,-2.8)node[right]{ $P_C(t)(l_C)$ for $T_{P_C}(\infty)$};
 \draw [->, >=stealth](6,0) parabola  (8.2,-2.0)node[above right]{\vdots};
 \draw [->, >=stealth](6,0) parabola  (8.2,-2.0)node[right]{\vdots};
\draw [->, >=stealth](6,0) parabola  (8.2,-1.5)node[above right]{\vdots};
\draw [->, >=stealth](6,0) parabola  (8.2,-1)node[above right]{\vdots};
\draw [->, >=stealth](6,0) parabola  (8.2,-0.5)node[above right]{\vdots};

\end{tikzpicture}

     Suppose there is a language $L_D$ and cardinality of $L_D$ is $\aleph_0$, and $l_D \in L_D$.There exist a Turing machine $P_D$ that can decide $L_D$, i.e., and $L(P_D)=L_D$ and $P_D$ halts on all inputs.We want to know the cardinality of $f(Z,T)$ transformed problems if we apply $f(Z,T)$ transform on $L_D$.We change $P_D$ into $P_D(t)$ and input is from $L_D$ through stipulation machine.If $|T_{P_D}|=\aleph_0$ then $f(Z,T)$ transform produces $\aleph_0 ^ {\aleph_0}$ number of combinations, and each combination is $f(Z,T)$ transformed language.There are uncountable number of $f(Z,T)$ transformed languages.

\begin{equation*}
    L_D=\{ l_{D(1)},\text{ } l_{D(2)},\text{ } l_{D(3)},\text{ } l_{D(4)},\text{ } l_{D(5)}\text{ } \hdots \text{ } \}
\end{equation*}

\hspace{-6mm}Suppose there is a language $L_E$ and cardinality of $L_E$ is $\aleph_0$, and we apply $f(Z,T)$ transform on $L_E$.If $n$ represents string number in $L_E$ and $0<T_{E} \leq \infty$ then Table $1$.represents $f(Z,T)$ transformed $L_E$.If $P_E$ is decider of $L_E$ and $l_E \in L_E$ then $P_E(t)(l_E)=P_E(l_E)$ for $T_E=\infty$ and $l_E(n\text{ } \infty)=L_E$.Therefore, it can be stated that the Language $L_E$ is subset of $f(Z,T)$ transformed $L_E$, and left most column of of following table represents language $L_E$.
\vspace{-4mm}
\begin{table}[h!]
    \centering
    \caption{ $f(Z,T)$ Transformed Problems for Language $L_E$}
    \renewcommand{\arraystretch}{1.2} % Increase cell size
   $ \begin{array}{|c|c|c|c|c|c|c|c|c|c|}
        \hline
        \mathrm{n} \backslash \mathrm{T_E} & 1 & 2 & 3 & 4 & 5 & 6 & 7 & \ldots & \infty \\
        \hline
        1 & l_E(1,1) & l_E(1,2) & l_E(1,3) & l_E(1,4) & l_E(1,5) & l_E(1,6) & l_E(1,7) & \ldots & l_E(1, \infty) \\
        \hline
        2 & l_E(2,1) & l_E(2,2) & l_E(2,3) & l_E(2,4) & l_E(2,5) & l_E(2,6) & l_E(2,7) & \ldots & l_E(2, \infty) \\
        \hline
        3 & l_E(3,1) & l_E(3,2) & l_E(3,3) & l_E(3,4) & l_E(3,5) & l_E(3,6) & l_E(3,7) & \ldots & l_E(3, \infty) \\
        \hline
        4 & l_E(4,1) & l_E(4,2) & l_E(4,3) & l_E(4,4) & l_E(4,5) & l_E(4,6) & l_E(4,7) & \ldots & l_E(4, \infty) \\
        \hline
        5 & l_E(5,1) & l_E(5,2) & l_E(5,3) & l_E(5,4) & l_E(5,5) & l_E(5,6) & l_E(5,7) & \ldots & l_E(5, \infty) \\
        \hline
        6 & l_E(6,1) & l_E(6,2) & l_E(6,3) & l_E(6,4) & l_E(6,5) & l_E(6,6) & l_E(6,7) & \ldots & l_E(6, \infty) \\
        \hline
        7 & l_E(7,1) & l_E(7,2) & l_E(7,3) & l_E(7,4) & l_E(7,5) & l_E(7,6) & l_E(7,7) & \ldots & l_E(7, \infty) \\
        \hline
        \vdots & \vdots & \vdots & \vdots & \vdots & \vdots & \vdots & \vdots & \ddots & \vdots \\
        \hline
        \infty & l_E(\infty, 1) & l_E(\infty, 2) & l_E(\infty, 3) & l_E(\infty, 4) & l_E(\infty, 5) & l_E(\infty, 6) & l_E(\infty, 7) & \ldots & l_E(\infty, \infty) \\
        \hline
    \end{array}$
\end{table}
%%%%%%%%%%%%%%%%%%%%%%%%%  Table Above %%%%%%%%%%%%%%%%%%%%%%%%%%
\\Cardinality of Turing machines is countably infinite and cardinality of $f(Z,T)$ transformed languages is uncountable, and this difference of cardinalities implies the existence of $f(Z,T)$ transformed problems that are uncomputable by Turing machines.

\section{Mathematical Formulations of $f(Z,T)$ Transform}

During construction of $f(Z,T)$ transform we assumed that $W_P = 1 - 2^{-T_P}$ in (3). But if we assume that $W_P = 1 - 2^{-\sfrac{T_P}{c}}$ and $c$ is some positive constant then we get (34 and 35) after replacing $1 - 2^{-T_P}$ with $1 - 2^{-\sfrac{T_P}{c}}$ in (6 and 32).
\begin{equation}
    f(P,T) = \frac{N_P T_t}{1 - 2^{-\sfrac{T_P}{c}}}
\end{equation}
\begin{equation}
   f_n(P,T) = \frac{N_P T_t}{T_P}
\end{equation}
If constant $c=N_P T_t$ and we replace the value of $c$ with $N_P T_t$ in (34) then we get (36).We can describe computability of every $f(Z,T)$ Transformed problem through (35 and 36).There are four possible outcomes when we try to compute any $f(Z,T)$ transformed problems by a Turing machine (or by a Turing complete model of computation).
\begin{equation}
    f(P,T) = \frac{N_P T_t}{1 - 2^{-\left(\sfrac{T_P}{N_P T_t}\right)}}
\end{equation}

\begin{equation*}
\hspace{-43mm} \textbf{Case I.} \hspace{40mm} \text{ If  } f_n(P,T)=0 \text{   then } f(P,T)=f(P)
\end{equation*}
\begin{equation*}
  \hspace{-25mm} \textbf{Case II.} \hspace{35mm}  \text{ If  } 0< f_n(P,T)<1 \text{   then } f(P) < f(P,T)<2f(P)
\end{equation*}
\begin{equation*}
 \hspace{-38mm} \textbf{Case III.} \hspace{40mm}   \text{ If  } f_n(P,T)=1 \text{   then } f(P,T)=2f(P)
\end{equation*}
\begin{equation*}
\hspace{-38mm} \textbf{Case IV.} \hspace{40mm}    \text{ If  } f_n(P,T)>1 \text{   then }  f(P,T)>2f(P)
\end{equation*}
We apply $f(Z,T)$ transform only on computable problems, and transformed problems can be computable (case I and II) or uncomputable (case IV), and sometime the computability of the transformed problem is highly sensitive toward choice of computational scheme(case III).If $T_P=\sfrac{1}{F_P}$ and $T_t=\sfrac{1}{F_t}$ then we can represent normalisation of transformed problem through frequencies in following way.

\begin{equation*}
    f_n(P, F) = \frac{N_P F_P}{F_t}
\end{equation*}
There are two frequencies in $f_n(P,F)$, i.e., $F_t$ and $F_P$.The frequency $F_t$ is related to clock speed of Turing machine and $F_P$ describes the frequency of recurring inputs.If frequency of recurring inputs is higher than certain threshold frequency then computable problem turns into uncomputable problem.It is helpful to describe normalization through frequencies if transformed problems involve measuring some frequencies.%%%%%%%%%%%%%%%%%%%%%%%%%%%%%%%%%%%%%%%%%%%%%%%%%%%%%%%%%%%%%%%%%%%%%%

\section{Summary}
~~~~In this paper we develop a transform technique and apply it on some decidable language, and show that cardinality of transformed languages is uncountable.There are countably infinite Turing machines but uncountable number of transformed languages, and this difference in cardinality implies the existence of transformed problems that are not computable.Furthermore, we present a realizable computational scheme that can demonstrate $f(Z,T)$ transform, and this scheme consists of Turing machine and Stipulation machine.Through $f(Z,T)$ transform, quite unexpectedly, we will be able to formalize some important natural phenomena.We will present experimental validation of this transform in future work, and those experiments will be designed and executed in such a way that it shall be easy to reproduce them.

\newpage

\bibliographystyle{unsrt}
\bibliography{pap5.bib}
\begin{flushright}
{\Large II}
\end{flushright}
\end{document}